\documentclass[aps,prb,showpacs,twocolumn, showkeys]{revtex4-1}
\usepackage{graphicx}

\newcommand*{\mb}{\mathbf}

\begin{document}

\title{Hybridization of quantum plasmon modes in coupled nanowires: From the classical to the tunneling regime}

\author{Kirsten Andersen}
\email{kiran@fysik.dtu.dk}
\affiliation{Center for Atomic-scale Materials Design, Department of
Physics \\ Technical University of Denmark, DK - 2800 Kgs. Lyngby, Denmark}

\author{Kristian L. Jensen}
\affiliation{Center for Atomic-scale Materials Design, Department of
Physics \\ Technical University of Denmark, DK - 2800 Kgs. Lyngby, Denmark}

\author{N. Asger Mortensen}
\affiliation{Department of
Photonics Engineering, and Center for Nanostructured Graphene \\ Technical University of Denmark, DK - 2800 Kgs. Lyngby, Denmark}

\author{Kristian S. Thygesen}
\email{thygesen@fysik.dtu.dk}
\affiliation{Center for Atomic-scale Materials Design, Department of
Physics,  and Center for Nanostructured Graphene  \\ Technical University of Denmark, DK - 2800 Kgs. Lyngby, Denmark}

\date{\today}

\begin{abstract}
  We present full quantum mechanical calculations of the hybridized
  plasmon modes of two nanowires at small separation, providing real
  space visualization of the modes in the transition from the
  classical to the quantum tunneling regime.  The plasmon modes are
  obtained as certain eigenfunctions of the dynamical dielectric
  function which is computed using time dependent density functional
  theory (TDDFT).  For freestanding wires, the energy of both surface
  and bulk plasmon modes deviate from the classical result for low
  wire radii and high momentum transfer due to effects of electron
  spill-out, non-local response, and coupling to single-particle
  transitions. For the wire dimer the shape of the hybridized plasmon modes are continuously altered with decreasing separation, and below 6 \AA\ the energy dispersion of the modes deviate from classical results due to the onset of weak tunneling. 
 Below 2-3 \AA\ separation this mode is replaced by a charge-transfer plasmon which blue shifts with decreasing separation in agreement with experiment, and marks the onset of the strong tunneling regime. 
\end{abstract}


\maketitle

\section{Introduction}
Plasmons are collective oscillatory modes of the electron system that are
found in solids, at extended surfaces, and at the surface of confined
metal structures. \cite{Pitarke_2007}  The localized surface plasmon
resonances sustained by metallic nanoparticles \cite{Willets_2007}
have found widespread use in applications ranging from molecular spectroscopy \cite{Kneipp_1997} and chemical sensing, \cite{Larsson_2009} to
photocatalysis \cite{Hirakawa_2005} and photovoltaics.
\cite{Atwater_2010} Applications of plasmons in next generation
nano-electronics\cite{Macdonald_2008} and quantum information
technology \cite{chang2006_quantum, Huck_2011} have also been envisioned.  Advances in
nano-fabrication techniques and transmission electron microscopy have
recently enabled the study of plasmon excitations on a subnanometer
length scale.\cite{Abajo_2010} In particular, the use of highly
confined electron beams can provide spatial resolution of the electron
energy loss spectrum (EELS) yielding unprecedented information about
the dynamical properties of confined electron systems. \cite{bosman2007mapping, koh_2009, nicoletti2011_surface, guiton2011correlated, Alber_2012}

When the size of a plasmonic structure reaches the nano-scale, quantum
effects begin to play a dominant role and the widely used classical models of electronic response become insufficient. \cite{Scholl_2012, raza2013_blueshift}  In this size regime, the theoretical description must be refined to account for
quantum effects such as surface scattering, electron spill-out at the
surface, coupling of plasmons to single-particle excitations, and the
non-locality of the electronic response. We note that some of these effects, in particular the non-local response and spill-out, can be incorporated into semi-classical hydrodynamic models.
\cite{Raza_2011, David_2011,Pitarke_2007} However, the validity and predictive power of such an approach in the true quantum regime is not fully clarified, and the standard hydrodynamic approximation, not including spill-out, has been found to fail for calculations on small nanometric gabs due to a insufficient description of induced charges in the interface. \cite{stella2013_performance}

One important hallmark of the quantum regime is electron tunneling 
which occurs when two systems are brought into close proximity. In fact tunneling leads to a
break-down of the local field enhancement predicted by the classical theory at sub-nanometer separations,
and leads to the formation 
of a charge-transfer plasmon.\cite{Romero_2006,
  Lassiter_2008, Zuloaga_2009, Esteban_2012} Classically, the plasmon
modes of a composite system can be obtained as even/odd combinations of the plasmon modes of the individual sub-systems. \cite{Prodan_2003, Wang_2006} 

Recent experiments by Savage et al.~\cite{Savage_2012} on coupled nanoparticles at finely controlled separation, found that the onset of the quantum regime is marked by a beginning blue shift of the hybridized modes, followed by the entrance of charge-transfer modes at conductive contact. Through the use of a quantum corrected model, they predict the onset of quantum tunneling effects at around 0.3 nm particle separation. This conclusion is supported by experiments by Scholl et al.~\cite{scholl2012observation} on silver nanoparticles, where the bonding dipole plasmon was replaced by a charge-transfer plasmon below 0.27 nm. Other experiments on film-coupled nanoparticles \cite{ciraci2012_probing} found the energy of hybridized modes to be blue shifted relative to the classical result at separations less than 1 nm, a similar blueshifting anticipated for nanowire dimers within a hydrodynamic description. \cite{toscano2012modified}  However, until now the influence of quantum effects on the spatial shape of the hybridized modes has not been explored.

In this work we apply a recently developed method \cite{Andersen_2012} to compute and visualize the hybridization of the quantum plasmons of two parallel metallic nanowires at separations between 0 and 2 nm. The evolution of the spatial form of the hybrid modes and their energies is used to determine the regimes where quantum effects are of importance.
We first show that the plasmon modes of single wires can be classified
as surface or bulk modes with a specific angular symmetry index $m$
and radial node index $n$. The maximum number of modes found for a
given wire radius is limited by Landau damping which increases with
$n$ and $m$. For the double wire system, we find several hybridized plasmon
modes. The induced potential of the hybrid modes changes in  shape in a continuous manner with decreasing separation and below 6 \AA\ the energies deviate from the classical result, in agreement with the observed blue-shift of the bonding dipole plasmon with decreasing dimer separation. We explain this by a beginning weak tunneling below 6 \AA\ where the electron densities start to overlap. At separations below 2-3 \AA, the onset depending slightly on wire diameter, we enter the regime of strong tunneling where the bonding dipole plasmon is quenched and replaced by a charge-transfer mode.
\section{Method}
\subsection{Theory of plasmon eigenmodes}
The plasmonic properties of a given material is contained in its frequency-dependent dielectric function, which relates the total and the externally applied potential to linear order:
\begin{equation}
\phi_{\mathrm{ext}}(r,\omega)= \int \epsilon(r,r',\omega) \phi_\mathrm{tot}(r',\omega)dr'.
\end{equation}
A self-sustained charge density oscillation, $\rho(r,\omega)$, can exist in a material if the related potential, satisfying the Poisson equation $\nabla^2 \phi = -4\pi \rho$, obeys the equation
\begin{equation}
 \int \epsilon(r,r',\omega)\phi(r',\omega)dr' = 0.
\end{equation}
This corresponds to the case of having a finite induced potential in the absence of an external potential, which is the criterion for the existence of a plasmon eigenmode. In general, this equation cannot be exactly satisfied due to a finite imaginary part originating from single-particle transitions which will lead to damping of the charge oscillation. When the damping in the material is small it is sufficient to require only that the real part of $\epsilon$ vanishes and use the following definition for the potential associated with a plasmon mode,
\begin{equation} \label{eq.2}
 \int \epsilon(r,r',\omega_n)\phi_n(r',\omega_n)dr' = i\Gamma_n \phi_n(r,\omega_n),
\end{equation} 
where $\Gamma_n$ is a real number. The plasmon modes are thus the eigenfunctions corresponding to purely imaginary eigenvalues of the dielectric function. Physically, they represent the potential associated with self-sustained charge-density oscillations damped by electron-hole pair formations at the rate $\Gamma_n$. The righthand-side eigenfunctions, $\phi_n(r',\omega_n)$, defines the induced potential of the eigenmodes, and the corresponding induced density $\rho_n(r',\omega_n)$ can be obtained as the lefthand-side (dual) eigenfunctions of the dielectric function. \cite{Andersen_2012} In the case of larger and frequency-dependent damping, a more accurate approach is to use the eigenvalues of $\epsilon^{-1}(r,r',\omega)$, and use the criterion:
\begin{equation}\label{eq.max}
\text{Im}\epsilon_n(\omega_n)^{-1}\text{ is a local maximum,}
\end{equation}
since a finite imaginary part will shift the peaks in the loss spectrum away from the zeros of $\text{Re}\epsilon_n(\omega)$. 

\begin{figure}
  \includegraphics[width = 0.48\textwidth]{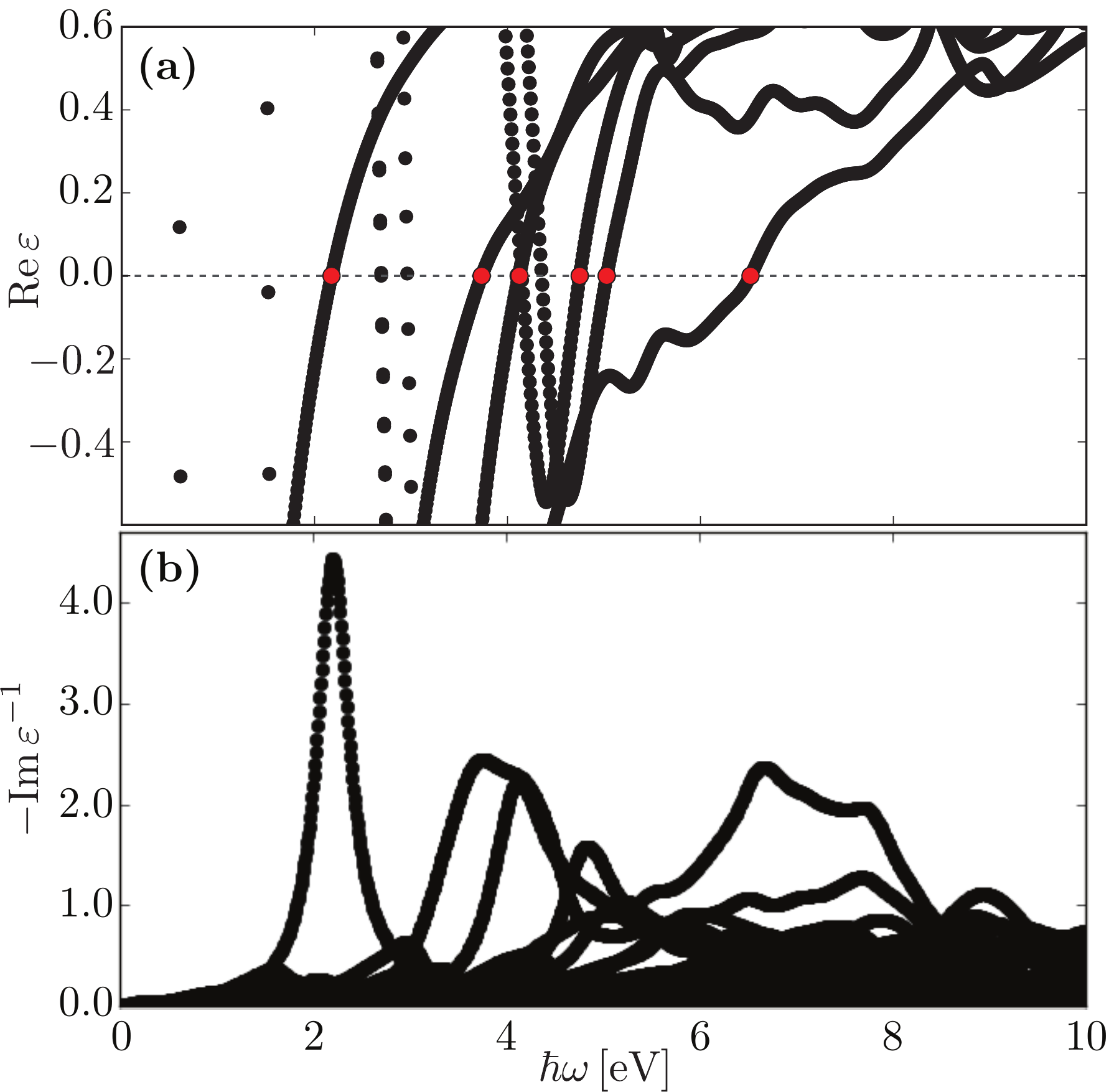}
  \caption{(a) Real part of the eigenvalues of the dielectric function $\epsilon_q(\omega, r,r')$ for a 5 
	\AA\ radius jellium wire. Each eigenvalue-curve that extends to negative values defines the existence of a
	plasmon eigenmode, and the crossing with zero from below (red points), approximates the plasmon energy of the current mode. (b) For a more correct description of the mode energies in the presence of large damping, maxima in the eigenvalues of $- \mathrm{Im} \, \epsilon^{-1}_q(r,r',\omega)$ are used to define the plasmon frequencies,  $\hbar \omega_n$.}	  \label{eigenvalues}
\end{figure}

In practice the plasmon eigenmodes are obtained by diagonalizing the dynamical dielectric function for each point on a frequency grid. This approach builds upon the concept of dielectric band-structure introduced by Baldereshi, \cite{baldereschi1979} but opposed to calculating the static dielectric band structure, this method enables us to identify plasmon modes and their energy. The dielectric eigenvalue spectrum results in a set of distinct eigenvalue-curves that evolves smoothly with energy, each curve corresponding to a separate plasmon eigenmode,  see Fig.~\ref{eigenvalues}. 
The eigenvalue-curves each resembles a Drude-Lorentz-like model for the dielectric function, and has been fitted to a single-pole model: 
\begin{equation}\label{eq3}
 \epsilon_n(\omega) = 1- \frac{\alpha_n}{\omega - \beta_n+ i \gamma_n},
\end{equation}
where only the real part is fitted, and used to find the mode strength, $\alpha_n$. The parameter $\beta_n$ corresponds to the average energy of the single particle transitions contributing the the given eigenmode, and $\gamma_n$ is the damping of the mode. 
\subsection{Computational approach}
\begin{figure*}
  \includegraphics[width = 0.65\textwidth]{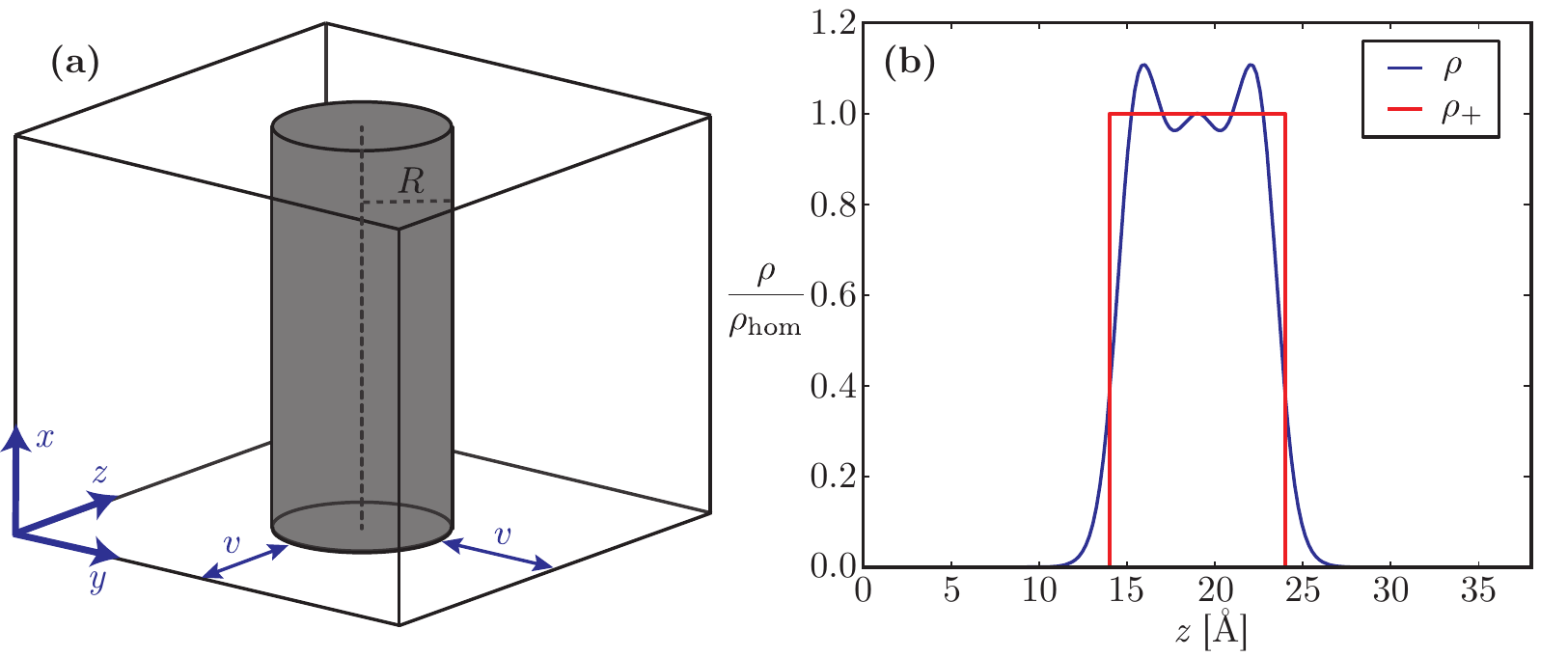}
  \caption{(a) Sketch of the unit cell with gray region indicating the positive charge density of a cylindrical jellium wire, surrounded by vacuum $2v = 28$ \AA. $R$ is varied from 1 to 8 \AA\ and the unit cell is 3 \AA\ along the wire (b) Cross-sectional view of the wire showing the fixed positive background density profile $\rho_+$ (red) and the converged electron density $\rho$ (blue), plotted relative to the free electron density of bulk Na, $\rho_{\mathrm{hom}}$. }
  \label{spillout}
\end{figure*}

The DFT calculations have been carried out with the electronic structure code GPAW \cite{GPAW1, GPAW2} using the Atomic Simulation Environment. \cite{ASE} We study metallic Na wires with radii in the range 2 to 8 \AA\,  modeled within the jellium approximation, that gives a good description of simple, free-electron like metals such as Na. The atomic nuclei of the nanowire are modeled with a constant positive background density that terminates abrupt at the wire edges, while the electron density is converged self-consistently, see Fig.~\ref{spillout}, showing the unit cell and the calculated densities within the jellium model. Electron spill-out extends approximately 3~\AA\ outside the jellium edge, and should be included for a correct description of surface plasmon modes in particular. This jellium approach thus includes the confinement of the electrons to the wire, and spill-out from the surface, but neglects the local field effects from the atomic nuclei. This allows us to study simple metals such as Na, where local field effects have been found to be negligible. \cite{Andersen_2012} We note that for the study of noble metals such as Ag and Au this jellium approach will not be justified since the $3d$ and $4d$ bands of these metals plays a large role in screening the plasmon excitations and therefore must be included. The local density approximation (LDA) was used for exchange-correlation. The unit cell was sampled on grid-points of 0.2 \AA\ spacing, and 20 k-points were used to map the 1D Brillouin zone along the wire, except for the study small momentum transfers in the response calculation where 80 k-points were used.  

The microscopic dielectric matrix, $\epsilon_{\mb{G},\mb{G'}}(\mb{q},\omega)$, was calculated in a plane-wave basis, using linear response TDDFT within the random phase approximation (RPA) \cite{Yan_2011}:
\begin{equation}
 \epsilon_{\mb{G},\mb{G}'}(\mb{q},\omega) = \delta_{\mb{G},\mb{G}'} - \frac{4\pi}{|\mb{q}+\mb{G}|^2} \chi^0_{\mb{G},\mb{G}'}(\mb{q},\omega),
\end{equation}
where the exchange correlation kernel $f_{xc}$ is neglected. The RPA level in general gives a good description of plasmon resonances but fails for the description of bound excitons. It performs well for the calculation of plasmon energies in simple metals, for example the surface plasmon energy of Mg(1000) is reproduced within 0.2 eV from the experimental value \cite{Yan_2011}. The RPA is exact in the high density limit, but is also expected to give a good description for simple metals with intermediate electron densities such as Na. The non-locality of the dielectric matrix is ensured by including off-diagonal matrix-elements in the reciprocal lattice vectors, $\mb{G}$ and $\mb{G'}$, where we used a 15 eV energy cutoff perpendicular to the wire.  Electronic states 10 eV above the Fermi level were included, and the dielectric function was sampled on an energy grid ranging from 0 to 10 eV with a spacing of 0.01 eV and a broadening of 0.2 eV. The wave-vector along the wire, $q_\parallel$, was varied between 0.03 $ \textrm{\AA}^{-1}$ and 0.8 $ \textrm{\AA}^{-1}$ in order to study the energy dispersion. Convergence have been checked with respect to all parameters, including the applied vacuum perpendicular to the wires, where a vacuum of 20 \AA\  ensures that there is no overlap of single-particle states from neighboring supercells. However, special care must be taken in the response calculation where the Coulomb kernel, $v(\mb{G}) = \frac{4 \pi}{| \mb{G}+\mb{q} |^2}$, diverges for $\mb{q}+\mb{G} \rightarrow 0$ and gives a long ranged potential for small $q$ that can not be compensated by adding vacuum in a realistic calculation. In order to avoid the overlap of induced potentials from neighboring supercells we use a truncated version of $v(\mb{G})$, that goes to zero as a step function at radial cutoff distance $R_{c}$ \cite{rozzi2006}:
\begin{widetext}
\begin{equation}
\tilde{v}^{1D} (G_\parallel,G_\perp) = \frac{4 \pi}{| G+q |^2} \biggl [ 1 + G_\perp R_{c} J_1(G_\perp R_{c})  K_0( | G_\parallel | R_{c})  -
 | G_\parallel | R_{c} J_0(G_\perp R) K_1(| G_\parallel | R_{c}) \biggr ],
\end{equation}
\end{widetext}
where $J$ and $K$ are the ordinary and modified cylindrical Bessel functions, and $G_\perp$ and $G_\parallel$ refer to components of the reciprocal lattice vectors perpendicular and parallel to the wire respectively. $R_{c}$ is set equal half of the width of the supercell, and this approach thus requires that the vacuum distance exceeds the diameter of the wire. 

The use of the truncated Coulomb potential was found to be important for a correct description of the plasmon mode energies for $q < 0.1 \AA^{-1}$, and in particular for modes with large macroscopic ($\mb{G}=0$) component. For the study of two coupled cylinders, we chose a $q$ that is sufficient large that the energy shift of the modes is less than 0.03 eV, and the truncation is not applied. 

\section{Results}
\subsection{Single wires}
\begin{figure*}
\includegraphics[width = 0.99\textwidth]{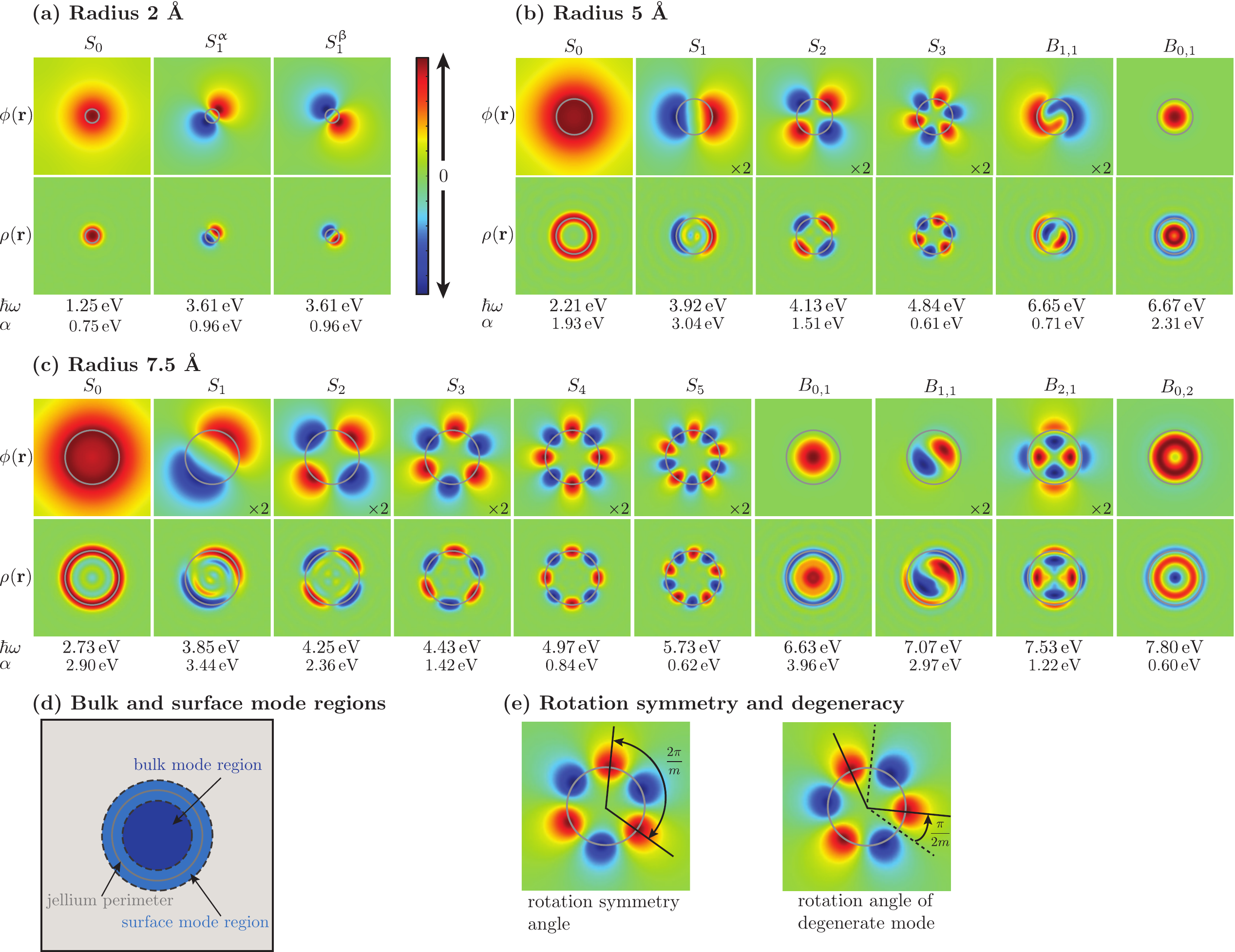}
 \caption{Plasmon eigenmodes of Na jellium wires of radii: (a) 2 \AA, (b) 5 \AA, and (c) 7.5 \AA, corresponding to induced potential, $\phi(\mb{r})$, and density, $\rho(\mb{r})$.  The strength, $\alpha$, and energy, $\hbar \omega$, is given for each mode. Modes with surface as well as bulk character are found, labelled S and B respectively, where the surface modes has a significant contribution in the spill-out region outside the jellium edge, see (d). The first subscript denotes the angular quantum number $m$, describing the $e^{i m \theta}$ rotational symmetry. The increase in $m$ leads to a larger effective wave-vector for the plasmon, $q_{\mathrm{eff}}$, resulting in an increase in energy with $m$. Also the Landau damping increases with $q_{\mathrm{eff}}$, which is why the a larger number of modes can be sustained by wires with larger radius. As explained in (e), modes with $m>0$ appear as degenerate pairs, rotated with $\pi/2m$ (plots labeled $\times 2$ when both are not shown).}
  \label{modes1}
\end{figure*}

The obtained plasmon eigenmodes for single wires of radius 2, 5 and 7.5 \AA\ is shown in Fig.~\ref{modes1}~(a)-(c), where the induced potential (first row) and density (second row) is plotted for increasing energy going left to right. The modes are characterized as surface, S, or bulk, B, determined by the relative weight outside and inside the jellium perimeter, sketched in Fig.~\ref{modes1} (d). The potential and density of surface modes has a large contribution in the spill-out region outside the jellium edge, and has lower energy than the bulk modes. The modes are labeled according to their angular quantum number $m$, accounting for the $e^{im\theta}$ angular symmetry, easily visualized for the surface modes as the number of standing waves along the wire perimeter. For the thinnest wire only modes up to $m=1$ are sustained, whereas modes with larger $m$ are found for increasing radii. This corresponds to a minimum plasmon wavelength, $\lambda_{\mathrm{min}} \approx 10 $ \AA\ that can be sustained by the surface, which is obtained by defining an effective wave-vector for the modes, $q_{\mathrm{eff}}^2 = q_\parallel^2 + q_\perp^2$. The parallel momentum transfer $q_\parallel$ is fixed in the calculation, and $q_\perp$ can be written: 
\begin{equation}\label{eq4}
q_\perp = \frac{2\pi}{\lambda_\perp} = \frac{m}{R},
\end{equation}
where $\lambda_\perp$ is the wavelength associated with the shape of modes perpendicular to the wire. When $m$ is sufficiently large, or $R$ small, the effective wave-vector becomes large enough for the mode to enter the regime of Landau damping (coupling to intra-band transitions with large $q$), until the point where the mode can no longer be sustained. This is also reflected in a mode strength, $\alpha$, that decreases with $m>1$ for the surface modes. 

For the bulk modes an additional quantum number, $n$, is required to account for radial dependence, where $B_{0,2}$ has an extra node in the induced density compared to $B_{0,1}$. From a group-theoretical analysis of the possible symmetries within the point-group $D_{\infty h}$, the modes with $m>0$ are expected to appear in degenerate pairs. This is exactly found for the calculated modes, where pairwise identical eigenvalue-curves results in identical modes with a relative rotation of $\pi/2m$, see Fig.~\ref{modes1} (e). 
\begin{figure*}
  \includegraphics[width = 0.99\textwidth]{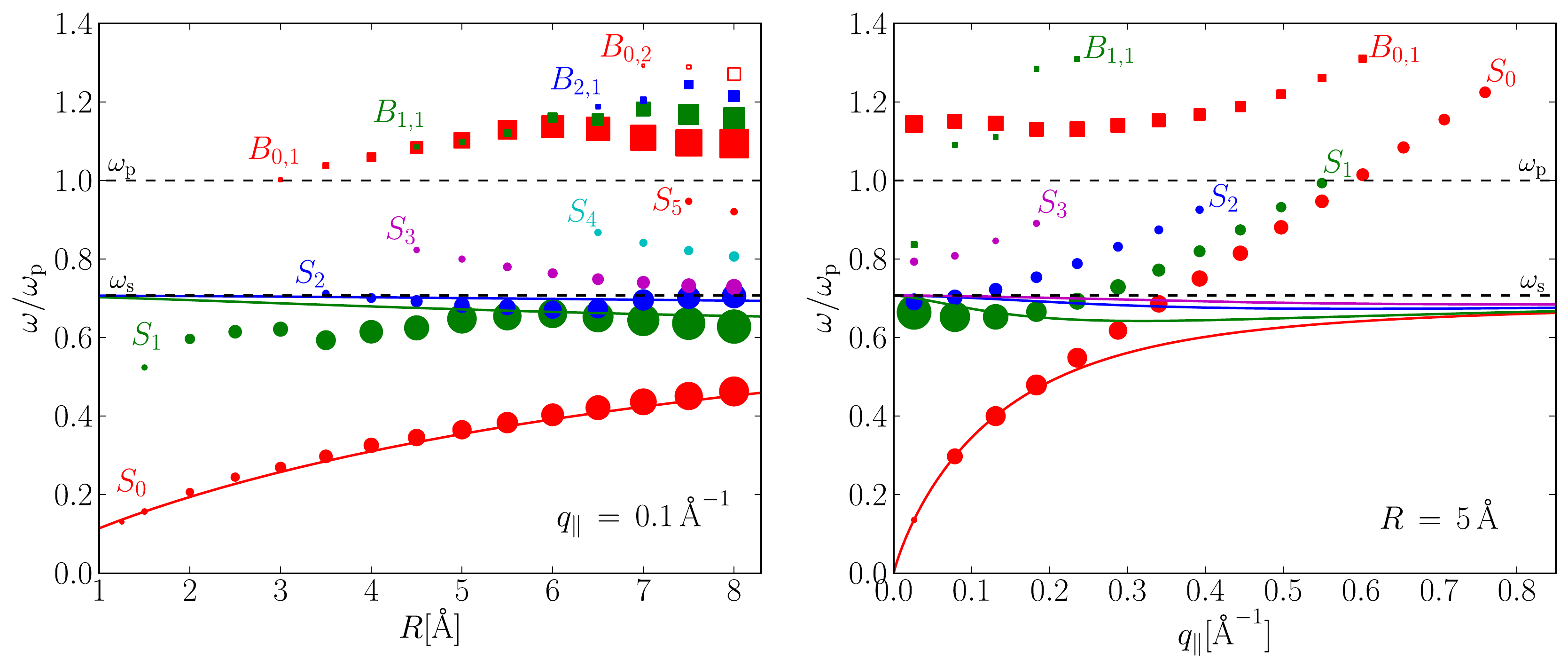}
  \caption{ Energy dispersion of plasmon eigenmodes for the quantum calculations (circles and squares) and the classical results for the surface modes (solid lines). The energy is plotted against: (a) Wire radius for fixed $q_\parallel = 0.10 \textrm{\AA}^{-1}$. (b) Wave-vector $q_\parallel$ for fixed $R = 5 \textrm{\AA}$. Circles and squares represent surface and bulk modes respectively, and the diameter/height of the points indicate the mode strength, $\alpha$, obtained by fitting the eigenvalues to a single pole model, Eq. \ref{eq3}. The $S_0$ mode, which has acoustic character, tends to zero energy for vanishing $R$ or $q$, where there is a good agreement with the classical result. For large $q$ the quantum model produces a positive dispersion for all modes, which is not captured by local models. For surface modes with $m>0$, there is an overall poor agreement due to effects from a larger $q_{\textrm{eff}}$ and spill-out.} 
   \label{dispersion1}
\end{figure*}
We have investigated the energy dispersion of the modes with $R$ and $q_\parallel$, shown in Fig.\ref{dispersion1}, where the result from our quantum calculation has been plotted together with the classical result obtained from solving Laplace's equation and using a local Drude model for the dielectric function. \cite{Gervasoni_2003} The classical model yields a dispersion relation for the surface modes that depends on a single dimensionless parameter, $q R$, where $q$ is the wave-vector parallel to the wire: 
\begin{equation}\label{classicaldisp}
\omega_m^2 = \omega_p^2 \; q R \; I^{\prime}_m(q R) \; K_m(q R),
\end{equation}
where $I_m$ is the modified Bessel's functions of order $m$, and $\omega_p$ is the bulk plasmon frequency. The acoustic character of the $S_0$ mode leads to an energy that goes to zero as a function of $q_\parallel R$, whereas the $S_{m>0}$ modes tends towards the classical surface plasmon frequency $\omega_P/\sqrt{2}$ for low $q_\parallel R$. For large $q_\parallel R$ all modes tends towards $\omega_P/\sqrt{2}$.

 When comparing to our quantum results, a very good agreement is seen for the $S_0$ mode at low $q_\parallel$, which breaks down for larger $q_\parallel$. Since the Drude model assumes the electronic response to be local in space, it cannot capture the dispersion with $q$ which explains the significant deviation for large $q_\parallel$, where the quantum result for the energy continue to increase. This $q$-dependence explains why the energy of the surface modes increases with larger $m$ (due to a larger $q_{\textrm{eff}}$), opposed to the classical result where the energy does not exceed $\omega_{\mathrm{P}}/\sqrt{2}$. 
This deviation from the classical result is also observed in semiclassical hydrodynamic models for waveguiding on nanowires. \cite{toscano2012nonlocal} However, $S_1$ and $S_2$, are seen to be red-shifted with respect to the classical value in the limit $q \rightarrow 0$, and for $S_1$ the redshift increases further with decreasing $R$.  A redshift is expected for decreasing size of simple metal nanostructures due to electron spill-out; as the surface to bulk ratio increases, the spill-out will lead to a decrease in the mean electron density, which results in a lower plasmon frequency. This redshift was also observed for the plasmon eigenmodes of thin slabs of sodium, \cite{Andersen_2012} and for quantum calculation on small Na clusters. \cite{Liebsch_1993} This is opposed to the behavior of silver nanostructures, where coupling to single-particle transitions originating from the 3d-states leads to blueshift with decreasing size. \cite{Liebsch_1993, Scholl_2012} Thus, there are two competing factors affecting the energy of the $S_{m>0}$ modes for decreasing R; the increase due to a larger perpendicular $q$, and the decrease due to spill-out, which might explain the oscillatory behavior of the $S_1$ mode for small $R$. 

When considering the strength of the modes (represented by the width of the labels in Fig. \ref{dispersion1}), there is a clear tendency for weakening of modes for small $R$ and large $q_\parallel$. This is due to a smaller number of electronic bands crossing the Fermi level for the thin wires, which gives a smaller number of transitions contributing to the plasmon in this limit. The weakening with increasing $q$ can be explained by increasing Landau damping as discussed previously. 

Even though wires modeled in this work are thin compared to a realistic setup, the results allow us to make some generalizations to thicker wires. For larger $R$ the number of surface modes $S_m$ are expected to increase according to the wire perimeter and the a minimum plasmon wavelength, 
$\lambda_{\mathrm{min}} \approx 10$ \AA, such that $m_\mathrm{max} = 2 \pi R / \lambda_{\mathrm{min}}$.  These higher order modes will not couple to optical fields, i.e. they are so called dark modes, but should be observable in a scanning EELS experiment. As seen from the energy dispersion in Fig.~\ref{dispersion1}, the energies for the surface modes approach the result of the classical model for increasing $R$, however the energy of the $S_{m=1}$ mode is still noticeably redshifted compared to the classical value at $R=8$ \AA. Thus, for this mode the impact of spill-out exceeds the size quantization effects, which lead to an overall redshift in energy.\cite{monreal2013_plasmons} For calculations on thin Na slabs, \cite{Andersen_2012} a redshift of the dipole plasmon mode was observed at 5 nm film thickness, and since spill-out will have a larger impact on 1D structures this indicates that the redshift should be observable for much thicker wires. The magnitude of the $q$-dispersion is expected to be independent on $R$ for large $q$, and interestingly we find an apparent linear dispersion of the surface modes with $q$ in this range. 
\begin{figure}
  \includegraphics[width = 0.40\textwidth]{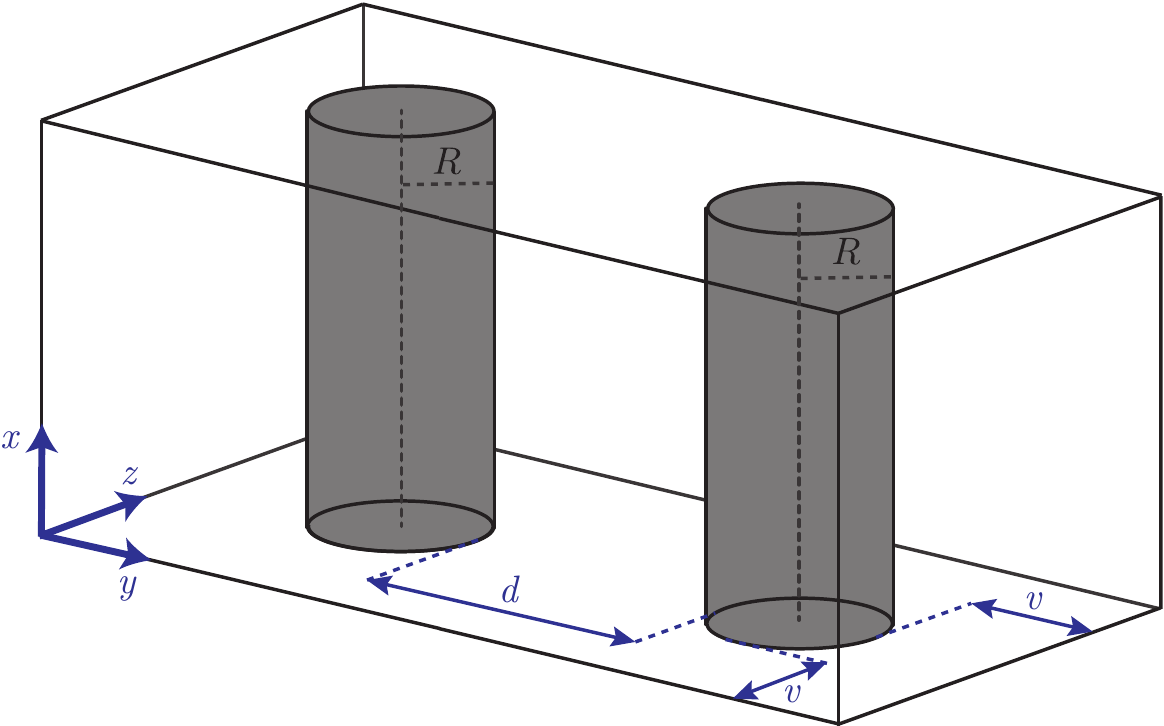}
  \caption{The unit cell containing two jellium wires of distance $d$. We consider separations in the range 0 to 20 \AA, and choose $R$ = 2.5 \AA\ for both wires, in order to study the case where only three modes exist for the individual wires.}
   \label{unit2}
\end{figure}

\subsection{Coupled nanowires}
\begin{figure*}
 \center
  \includegraphics[width = 1.01\textwidth]{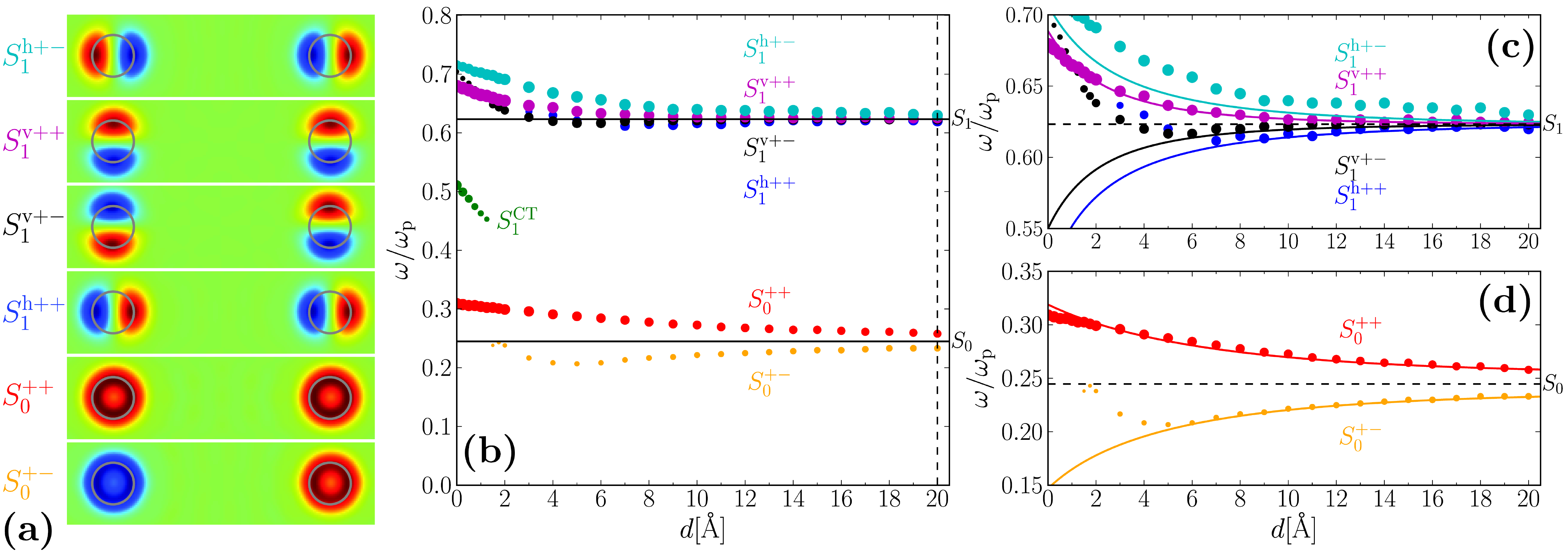}
  \caption{(a) Induced density of hybridized modes of two parallel wires with $R= 2.5$ \AA\ at 20 \AA\ separation. The modes can be identified as hybridization between the $S_0$, $S^\alpha_1$ and $S^\beta_1$ modes of the individual wires, where only modes of the same type couple due to symmetry restrictions. 
(b) The mode energies as a function of wire separation, $d$, with the mode strength indicated by the diameter of the points. The hybridization leads to a split in energy of the even, $++$, and odd, $+-$ mode-combinations, plotted together with the classical result (solid curves) in (c) and (d). 
At 2 \AA\ separation the $S_0^{+-}$ and $S_1^{\textrm{h}++}$ modes are quenched out due to tunneling and a charge-transfer mode $S_1^{\mathrm{CT}}$ appears.}
   \label{disp2}
\end{figure*}

After establishing an understanding of the quantum modes of single wires, we are ready to study the hybridization of modes between two parallel wires of small separation. The unit cell is sketched in Fig.~\ref{unit2}, where the separation, $d$, is varied from 20 \AA\ down to zero measured from the jellium edge, and the amount of vacuum separating the wires of neighboring super-cells are kept constant at 20 \AA. The hybridization of plasmon modes is not unlike the formation of molecular orbitals as bonding and antibonding combinations of atomic orbitals, \cite{Prodan_2003,Wang_2006} where the new hybridized modes correspond to the even and odd combinations of isolated modes of the individual structures. At very small separations however, this simplified description no longer holds since overlap of the electronic states and eventual tunneling will alter the resonances. We have applied our quantum approach to provide real-space visualization of the hybridized modes in the transition from the non overlap to the overlap regime.

\begin{figure*}
  \includegraphics[width = 0.99\textwidth]{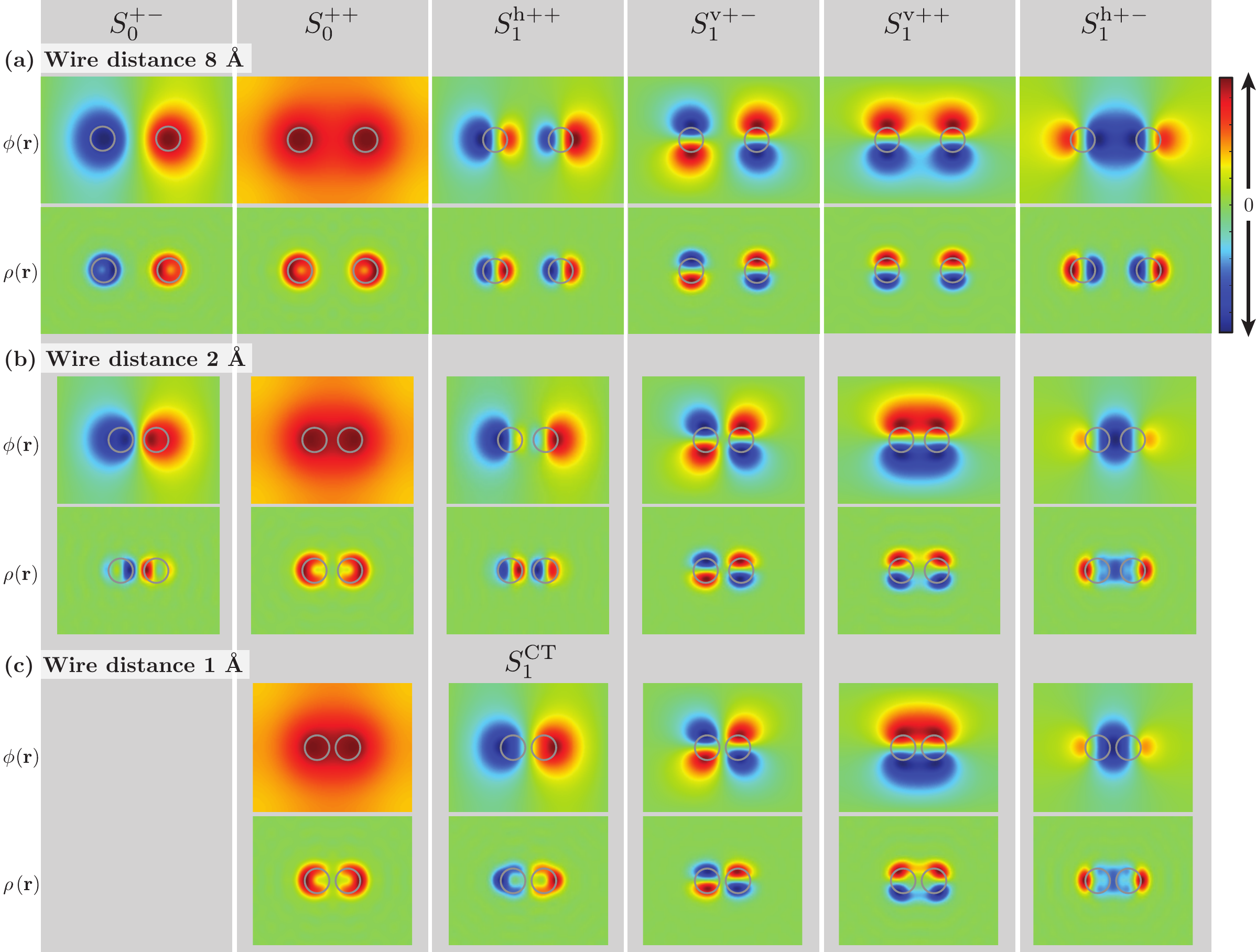}
  \caption{The plasmon eigenmodes for two parallel nanowires with $R = 2.5$ \AA\ for separations (a) 8 \AA, (b) 2 \AA\ and (c) 1 \AA.  Already at 8 \AA\ separation the potential, $\phi(r)$, and density, $\rho(r)$, of the modes are slightly modified, the change being most dominant for the $S_1^{\textrm{h}++}$ and $S_0^{+-}$ modes. At 2 \AA\ these mode are almost quenched out due to tunneling, and the induced density is shifted into the bulk. Below 2 \AA\ a new charge-transfer mode $S_1^{\mathrm{CT}}$, where the density oscillates between the wires, appears. For the other modes there is a continuous evolution in shape from the non-touching to the touching regime, as seen for the mode energies as well. }
   \label{modes2}
\end{figure*}

 In order to limit the number of hybridized modes we choose both wire radii equal to 2.5 \AA, where only the $S_0$ and the two degenerate $S_1$ modes exists. Though the wires used in this study is thinner than a realistic experimental setup, it allow us to perform a qualitative study of the evolution of the hybridized modes with separation and give a quantitative discussion on the onset of quantum effects. 

 Only modes of similar symmetry with respect to the $y$-axis will couple to form hybridized modes; for example, an $S_0$ mode at one wire will be unable of inducing a $S_1$-like potential at a neighboring wire, and no self-sustained oscillation can occur. Therefore the modes only couple to those of the same type ($S_1$ couple to the $S_1$ mode of same rotation), and six hybridized modes is predicted to exists for this structure. 

Six modes are found in the quantum calculation as shown in Fig.~\ref{disp2}(a), with eigenmodes calculated for a 20 \AA\ separation of the wires. In this regime of zero overlap between the electronic states, the modes can be classified according to conventional hybridization theory as even or odd combinations of the initial modes, labelled $++$ and $+-$ respectively, where the four $S_1$-types are further divided into vertical, $\textrm{v}$, and horizontal, $\textrm{h}$, according to the direction of polarization. As can be seen in Fig.~\ref{modes2} showing the induced potential and density for 8 \AA, 2 \AA, and 1 \AA\ separations, this description is no longer valid for small separations. 

The energy dispersion of the modes as a function of $d$ is shown in Fig.~\ref{disp2}(b), where the $+-$ and $++$ combinations split in energy for intermediate separations. A red-shift is found for the bonding  modes where opposite charges face each other, i.e. $S_0^{+-}$, $S_1^{\textrm{h}++}$, and  $S_1^{\textrm{v}+-}$. This energy shift can be explained by the electrostatic attraction between induced charges of opposite sign that leads to a lowering of the plasmon frequency (and in reverse for the remaining anti-bonding modes). The resonance of largest technological importance is the $S_0^{+-}$ mode, also known as the bonding dipole plasmon, which has a large cross-section for coupling to optical fields.

The interaction between the wires increases for decreasing separations which leads to a larger splitting, until the weak tunneling regime is entered around 6 \AA. The shift is more distinct for the $S_0$-types that also has the most long-ranged potential that explains the stronger coupling and larger splitting of bonding and anti-bonding modes. The potential from the electric dipole of the $S_1$ modes decreases faster along $y$, in particular for the vertical modes, which again shows a smaller splitting than the horizontal modes.

In Fig.~\ref{disp2} (c) and (d) the quantum results have been plotted together with the results of a classical model. The later has been obtained by solving Laplace's equation for a system of two cylinders of equal radii $R$ and dielectric constant $\epsilon = 1- \frac{\omega_P^2}{\omega^2}$. Following a general approach by Moradi, \cite{Moradi_2011} we have derived the expressions for hybridized $S_0$ and $S_1$ modes for this system, see Appendix.  

Above the regime of weak tunneling there is a good agreement between the classical and quantum results for the energy splitting. Below $d\approx 6$ \AA\ however, the bonding modes, $S_0^{+-}$, $S_1^{\textrm{h}++}$, and  $S_1^{\textrm{v}+-}$, are blueshifted with respect to the classical model as can be explained by a reduction of the induced charges in the interface due to tunneling. 


Our approach for visualizing the eigenmodes from a quantum calculation provides a clear picture of how the plasmon modes are altered due to this quantum effect. As can be seen in Fig.~\ref{modes2}, already at 8 \AA\ separation some eigenmodes are modified compared to the large separation limit shown in Fig. \ref{disp2} (a). The modes are seen to be gradually altered with decreasing separation, with the $S_0^{+-}$ and $S_1^{\textrm{h}++}$ modes being most seriously affected due to the facing of opposite charges. The potential between the wires decreases in amplitude going from 8 \AA\ to 2 \AA\ separation, and the induced density is shifted further into the bulk. A decrease in mode strength (indicated by the marker diameter in Fig.~\ref{disp2}) is also found for these modes upon approaching 2 \AA\ separation. At 1 \AA\ both modes are quenched due to tunneling and have been replaced by a dipolar charge-transfer mode, here named $S_1^{\mathrm{CT}}$, and the induced density is now located at the other side of the wires, away from the interface. That this mode is different in nature than the $S_1^{\textrm{h}++}$ and $S_0^{+-}$ is also clear from Fig.~\ref{disp2}(b), where it appears as a separate line in the energy dispersion below 2 \AA, and increases in strength for smaller separations. In this touching regime with large degree of tunneling, the structure can be regarded as one (non-cylindrical) wire; however, for the other modes there is a continuous transition from the non-touching to the touching regime. The $S_1^{\textrm{v}++}$ and $S_1^{\textrm{h}+-}$ modes gradually build up induced density in the region between the wires, and $S_1^{\textrm{v}+-}$ ends up resembling a quadropolar ($m$=2) mode for the joint structure. 

We note that small thickness of nano-wires used in this study should be considered when making quantitative conclusions about the onset of quantum effects. For a dimer consisting of thicker wires realistic for experimental studies, the overlap of the electronic states could increase due to a lower surface curvature, which is expected to give rise to a earlier onset of quantum effects. For very thick wires the result should approach that of an bimetallic interface of two planar surfaces. Our corresponding calculations on such an interface, modeled by two infinite Na jellium films of varying separation, yielded the presence of a charge-transfer mode already at 3 \AA\ separation. This should be compared to the onset just below 2 \AA\ for the small dimer; Hence, for thicker wires the onset of the charge-transfer mode is expected to shift slightly to larger separations and approach that of a bimetallic interface. Quantum calculations on a corresponding Na nano-wire dimer or diameters up to $9.8$ \AA\ has recently been presented, \cite{teperik2013_quantum} that find an onset of the charge transfer plasmon close to 2 \AA\ separation measured from the jellium edge, in good agreement with our results.
  
Another factor that could potentially affect the onset of tunneling is the $r_s$ parameter defining the electron density in the jellium model. We note that choosing a smaller $r_s$ parameter than the one of Na will lead to a larger plasmon frequency due to the increase in density, but could also increase the workfunction of the surface and potentially shift the onset of tunneling to smaller separations. \cite{lang1970theory} However, our calculations on a dimer of Al jellium films did not change the onset of the charge-transfer plasmon compared to Na jellium films, so we do not expect this parameter to give large quantitative differences for simple metals.
 
\section{Conclusion}
We have given a visual demonstration of the plasmon eigenmodes in nano-wires and shown how the hybridized plasmon modes of a two-wire system is modified due to quantum effects for decreasing separation. For wire separations above $10$ \AA\  the modes are well characterized according to conventional hybridization theory, and a good correspondence was found with a classical theory for the mode energies until the onset of tunneling around $6$ \AA\ separation. Below this limit our quantum calculations produced a blueshift of the bonding plasmon modes with decreasing separation, due to the decrease in induced charge at the interface caused by tunneling. However; only for separations below 2-3 \AA\ was the tunneling sufficiently strong for a true charge-transfer plasmon to appear.
\appendix*
\section{Classical hybridization model for a two-wire system}
We have derived the classical expression for the energy of the hybridized modes of a system of two cylinders of equal radii $R$ of separation $d_c$ between the two centers, where the general approach by Moradi \cite{Moradi_2011} simplifies to finding the zeros of the determinant of the matrix: $(\mathbf{M}^2- I)$, where the elements of $\mathbf{M}$ are written:
\begin{equation}\label{classicaldispdouble}
M_{mn} = \frac{\omega_m^2}{\omega^2-\omega_m^2} \frac{K_{m+n}(q d_c) I_{m}(q R)}{K_{n}(q R)},
\end{equation}
where the $m,n$ indices corresponds to angular quantum numbers, and $\omega_m$ is the energy of the $S_m$ mode of a single wire given in Eq.~\ref{classicaldisp}. For the coupling of two $S_0$ modes both $m,n = 0$, and solving for omega yields two solutions: 
\begin{equation}\label{classicaldispdoubleS0}
\omega_{\pm}^ {2}= \omega_{0}^2 \biggl[1 \pm K_0(q d_c) \frac{I_0(q R)}{K_0(q R)}\biggr],
\end{equation}

accounting for the splitting of the $S_0^{+-}$ and $S_0^{++}$ modes. Due to the small discrepancy between the quantum results and the classical model for a single wire (due to dispersion with $q$ and $R$), we have set $\omega_0$ equal to the result from the quantum calculation for $S_0$ for a clearer comparison. 
For coupling of $S_1$ modes, $m,n = \pm 1$, and $(\mathbf{M}^2- I)$ is a 2 by 2 matrix, which leads to a total of four solutions for $\omega$:
\begin{widetext}
\begin{equation}\label{classicaldispdoubleS1}
\omega_{\{h,v\},\{\pm\}} ^2= \omega_{1}^2 \biggl[1 \{ \pm, \mp \} \frac{I_1(q R)}{K_1(q R)}\sqrt{K_2(q d_c)^2 + K_0(q d_c)^2 \pm 2 K_0(q d_c)K_2(q d_c) }  \;  \biggr],
\end{equation}
\end{widetext}
where $\omega_1$ has been set equal to the quantum result for $S_1$. 

\begin{acknowledgments}
KST acknowledges support from the Danish Council for Independent ResearchÕs Sapere Aude Program through grant no. 11-1051390. The Center for Nanostructured
Graphene CNG is sponsored by the Danish National Research Foundation, Project DNRF58.
\end{acknowledgments}

\bibliography{bibtex}

\end{document}